\documentclass{elsart1p}
\usepackage{graphicx}
\usepackage{amssymb}
\begin{document}
\begin{frontmatter}
%
%
%
\title{Electron Scattering off $^4$He with Three-Nucleon Forces}
%
%
\author{Sonia Bacca}
\address{TRIUMF, 4004 Wesbrook Mall,
Vancouver, B.C. V6J 2A3 , Canada}
\author{Nir Barnea}
\address{Racah Institute of Physics, Hebrew University, 91904, Jerusalem,
Israel}
\author{Winfried Leidemann, Giuseppina Orlandini}
\address{Dipartimento di Fisica, Universit\`{a} di Trento and INFN
(Gruppo Collegato di Trento), \\ via Sommarive 14, I-38100 Trento, Italy}
\begin{abstract}
An {\it ab initio} calculation of the $^4$He (e,e') inelastic longitudinal response function $R_L$ is presented.
Realistic two- and three-body forces are used. The four-body continuum dynamics is treated rigorously
with the help of the Lorentz integral transform.
The three-nucleon force reduces the quasi-elastic peak height by about 10\% for momentum transfers
$q$ between 300 and 500 MeV/c. Experimental data are well described, but not sufficiently precise to 
resolve this effect.
The reduction due to the three-nucleon force increases significantly at lower $q$ reaching up to about
40\% at $q=100$ MeV/c. However, at such $q$ values data are missing.

\end{abstract}
\end{frontmatter}
%
%
%
The longitudinal response function $R_L$ is given by
\begin{equation}
\label{frisp}
R_L(\omega,q)=\int \!\!\!\!\!\!\!\sum _{f}
|\left\langle \Psi_{f}| \hat\rho(q)|
\Psi _{0}\right\rangle|
^{2}\delta\left(E_{f}+\frac{q^2}{2M}-E_{0}-\omega \right)\,,
\end{equation}
where
$M$ is the target mass, $| \Psi_{0/f} \rangle$ and
$E_{0/f}$ denote the four-body initial and final state wave functions and energies,
respectively, while $\omega$ and $q$ are the energy and momentum
transfers. The charge density operator $\rho$ is defined as
\begin{equation}
{\hat\rho}(q)= \frac{e}{2} \sum_i \,(1 + \tau_i^3) \exp{[i {\bf q} \cdot {\bf r}_i]} \,,
\end{equation}
where $e$ is the proton charge and $\tau_i^3$ the isospin third component of nucleon $i$.
For a conventional calculation of $R_L$ one would need to know explicitly the four-body continuum state
wave functions $\Psi_f$. In the Lorentz integral transform (LIT) method~\cite{EFROS94} this difficulty is
circumvented  by considering instead of
$R_L(\omega,q)$ an integral transform
${\cal L}_L(\sigma,q)$ with a Lorentzian kernel defined for a complex
parameter $\sigma=\sigma_R+i\,\sigma_I$, which is then inverted in order to obtain $R_L(\omega,q)$ 
(see review \cite{REPORT07}). 

For the calculation of $R_L$ we take a realistic nuclear interaction consisting in the AV18 two-nucleon potential
\cite{AV18} and the UIX three-nucleon force (3NF) \cite{UIX}. The $^4$He ground-state wave function and the LIT are
calculated using expansions in hyperspherical harmonics with the EIHH \cite{EIHH,EIHH-3BF} and Lanczos \cite{Lanczos}
techniques (for more information concerning the calculation see \cite{BaB08}). 

\begin{figure}
\center
\resizebox*{9cm}{10cm}{\includegraphics[angle=0]{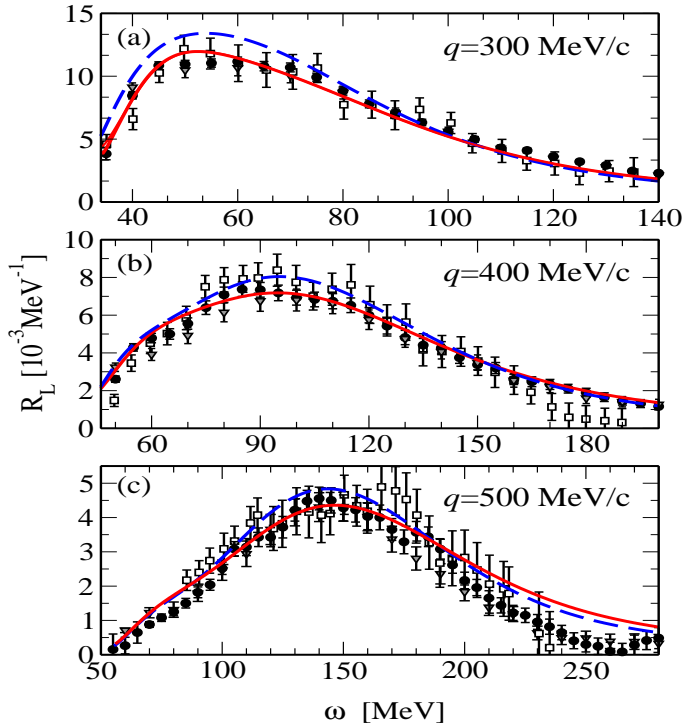}}
\caption{$R_L(\omega,q)$ at various $q$. Dashed lines: AV18; solid lines: AV18+UIX.
Data from Bates~\cite{Bates} (squares), Saclay~\cite{Saclay} (circles) and world data-set from \cite{Carlson}
(triangles).
}
\label{FIG2}
\end{figure}
In Fig.~1 we show $R_L$ at 300 MeV/c $\le$ $q$ $\le$ 500 MeV/c. One sees that the 3NF
reduces the quasi-elastic peak strength by about 10\%. Experimental data are described quite well 
by our full result, but they are not precise enough to resolve the 3NF effect. 
In Fig.~2 we illustrate
$R_L$ at lower $q$. One readily notes a very strong reduction at lower energies due to the 3NF,
which reaches up to about 40\%. The reduction cannot be attributed to a simple binding effect
as becomes evident from the also shown $R_L$ results with a semirealistic NN force (MT potential \cite{MT}). 
In fact, $^4$He binding energies are 24.3, 28.4, and 30.6 MeV for AV18, AV18+UIX, and MT potentials, respectively.
Even though the MT energy is closer to that of AV18+UIX, the MT $R_L$ is more similar to the $R_L$ 
of AV18 than to the AV18+UIX $R_L$.
At lower $q$ there is only one data set at 200 MeV/c \cite{Buki}, which is not sufficiently precise to draw
concrete conclusions.
\begin{figure}
\center
\resizebox*{8.cm}{7.cm}{\includegraphics[angle=0]{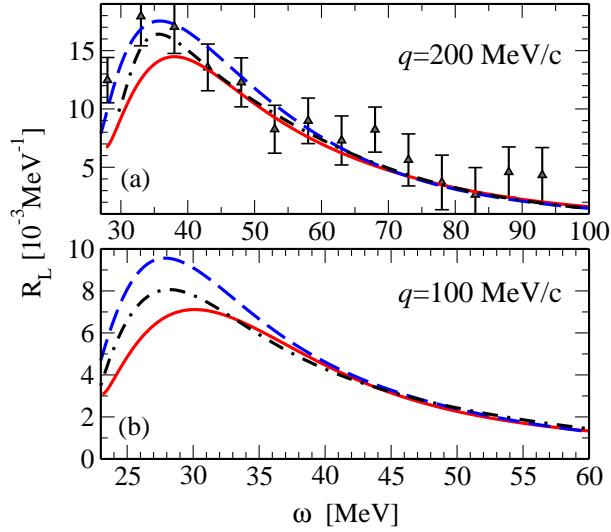}}
\caption{$R_L(\omega,q)$ at various $q$: AV18 (dashed), AV18+UIX (solid), MT (dash-dotted).
Data in (a) from~\cite{Buki}.
}
\label{FIG3}
\end{figure}

The work of S.B. was supported by the
Natural Sciences and Engineering Research Council (NSERC) and by the
National Research Council of Canada and the work of N.B. by the Israel Science Foundation (grant no.~361/05).
Numerical calculations were  performed at CINECA (Bologna).

%
%
%

%

\begin{thebibliography}{00}
%
%
%
%
\bibitem{EFROS94} V.~D. Efros, W. Leidemann, and G. Orlandini,
Phys. Lett. B {\bf 338}, 130 (1994).
\bibitem{REPORT07} V.~D. Efros, W. Leidemann,  G. Orlandini, and N. Barnea,
J. Phys. G: Nucl. Part. Phys. {\bf 34}, R459 (2007).
\bibitem{AV18} R.~B.~Wiringa, V.~G.~J.~Stoks, and R.~Schiavilla,
 Phys. Rev. C {\bf 51}, 38 (1995).
\bibitem{UIX} B.~S. Pudliner, V.~R. Pandharipande, J. Carlson, S.~C. Pieper, and
 R.~B. Wiringa, Phys. Rev. C {\bf 56}, 1720 (1997).
\bibitem{EIHH} N. Barnea, W. Leidemann, and G. Orlandini, 
Phys. Rev. C {\bf 61}, 054001 (2000); Nucl. Phys. {\bf A693}, 565 (2001).
\bibitem{EIHH-3BF} N. Barnea, V.~D. Efros, W. Leidemann, and G. Orlandini, Few-Body Syst.
{\bf 35}, 155 (2004).
\bibitem{Lanczos} M.~A. Marchisio, N. Barnea, W. Leidemann, and G. Orlandini, Few-Body Syst.
{\bf 33}, 259 (2003).
\bibitem{BaB08} S. Bacca, N. Barnea, W. Leidemann, and G. Orlandini, arXiv:0811.4624 [nucl-th].
\bibitem{Bates} S.~A. Dytman {\it et al.}, Phys. Rev. C {\bf 38}, 800 (1988).
\bibitem{Saclay} A. Zghiche {\it et al.}, Nucl.  Phys. {\bf A572}, 513 (1994).
\bibitem{Carlson} J. Carlson, J. Jourdan, R. Schiavilla, and I. Sick, Phys. Rev. C {\bf 65}, 024002 (2002).
\bibitem{MT} B.~S. Pudliner, V.~R. Pandharipande, J. Carlson, S.~C. Pieper, and
 R.~B. Wiringa, Phys. Rev. C {\bf 56}, 1720 (1997).
\bibitem{Buki} A.~Yu. Buki, I.~S. Timchenko, N.~G. Shevchemko, and I.~A. Nenko, Phys. Lett. B {\bf 641},
156 (2006).
%
\end{thebibliography}
\end{document}